\documentclass[fleqn,pra]{revtex4-2}

\usepackage{amsmath,graphicx}
\usepackage{verbatim}

\begin{document}

\title{Second-order corrections of orders $m\alpha^6$ and $m\alpha^6(m/M)$ to the spin-averaged energy in the HD$^+$ and H$_2^+$ ions.}

\author{Vladimir I. Korobov$^1$, Jean-Philippe Karr$^{2,3}$, and Zhen-Xiang Zhong$^4$}
\affiliation{$^1$Bogoliubov Laboratory of Theoretical Physics, Joint Institute for Nuclear Research, Dubna 141980, Russia}
\affiliation{$^2$Laboratoire Kastler Brossel, Sorbonne Universit\'e, CNRS, ENS-Universit\'e PSL, Coll\`ege de France, 4 place Jussieu, F-75005 Paris, France}
\affiliation{$^3$Universit\'e Evry-Paris-Saclay, Boulevard Fran\c cois Mitterrand, F-91000 Evry, France}
\affiliation{$^4$Center for Theoretical Physics, School of Physics and Optoelectronic Engineering, Hainan University, Haikou 570228, China}

\begin{abstract}
The relativistic second-order corrections at orders $m\alpha^6$ and $m\alpha^6(m/M)$ in hydrogen molecular ions are calculated. Convergence of numerical results is studied, which allows estimating the relative numerical uncertainty to be less then $10^{-5}$. This accuracy is sufficient to enable future improvement of theoretical predictions beyond the 1 ppt (part-per-trillion) precision level for the ro-vibrational transition frequencies.
\end{abstract}

\maketitle

Spectacular progress in laser spectroscopy of the hydrogen molecular ions HD$^+$~\cite{Alighanbari20,Patra20,Kortunov21,Schiller23} and H$_2^+$~\cite{Kienzler23,Schiller25} provides high potentiality for accurate determination of fundamental constants~\cite{Karr23,CODATA22, Schiller24,Karr25} such as the proton-to-electron mass ratio, Rydberg constant, and nuclear charge radii. It also allows to probe for manifestations of beyond-standard-model physics, in particular to obtain constraints on a hypothetical scalar fifth force between baryons~\cite{fifth_force21,Delaunay23}. Information on the spectrum of these ions has also been obtained from spectroscopy of Rydberg states of neutral hydrogen molecules, using subsequent extrapolation relying on the multi-channel quantum defect theory (see, e.g.,~\cite{Doran24a,Doran24b} and references therein). More extensive discussion on possible applications of hydrogen molecular ion physics may be found in \cite{SchillerCP}.

The current status of theory is as follows: the frequencies of vibrational transitions are calculated with a relative theoretical uncertainty of $u_{\rm th} \approx 8\!\times\!10^{-12}$~\cite{Korobov21}. Relativistic corrections of order $m\alpha^6$ have so far been calculated within the framework of the adiabatic approximation~\cite{Korobov07,Korobov08,Korobov17,Korobov21}, whereas recoil corrections were only estimated from the hydrogen atom theory \cite{Korobov17}. The first derivations of $m\alpha^6$-order corrections in a three-body formalism was performed in \cite{Zhong18,Korobov20}, but no numerical results have been published yet.

Here we present the first step toward more accurate determination of the $m\alpha^6$ and $m\alpha^6(m/M)$-order relativistic contribution to the ro-vibrational energies in the hydrogen molecular ions. More precisely, we calculate the complete set of second-order perturbative corrections that contribute to the spin-averaged transition energies in H$_2^+$ and HD$^+$ molecular ions.

Atomic units ($\hbar=e=m_e=1$) are employed throughout. We use the CODATA22 recommended values of fundamental constants \cite{CODATA22} in our calculations.

\section{Generalities}

We consider the general case of a one-electron diatomic molecule with nuclear charges $Z_1$, $Z_2$ and masses $M_1$, $M_2$. The positions of the nuclei and electron in the laboratory frame are denoted by $\mathbf{R}_a$ ($a=1,2$) and $\mathbf{r}_e$, and their impulses in the center-of-mass frame by $\mathbf{P}_a$ and $\mathbf{p}_e = -\mathbf{P}_1 -\mathbf{P}_2$. We use internal coordinates $\mathbf{r}_a = \mathbf{r}_e - \mathbf{R}_a$ and $\mathbf{R} = \mathbf{R}_2 - \mathbf{R}_1$. The nonrelativistic Hamiltonian in the center-of-mass frame is
\begin{equation}\label{Hamiltonian}
H_0 = \frac{\mathbf{P}_1^2}{2M_1}+\frac{\mathbf{P}_2^2}{2M_2}+\frac{\mathbf{p}_e^2}{2m}
      -\frac{Z_1}{r_1}-\frac{Z_2}{r_2}+\frac{Z_1Z_2}{R}
 = \frac{\mathbf{p}_1^2}{2\mu_{1}} + \frac{\mathbf{p}_2^2}{2\mu_2}+\frac{\mathbf{p}_1\mathbf{p}_2}{m} + V_1+V_2+V_{12},
\end{equation}
where $\mu_a^{-1}=M_a^{-1}\!+\!m^{-1}$, is the reduced mass of nucleus $a$, $\mathbf{p}_1=-\mathbf{P}_1$, and $\mathbf{p}_2=-\mathbf{P}_2$. We also use the notation $V_e=V_1\!+\!V_2$ for the electron potential energy.

The Breit-Pauli Hamiltonian at orders $m\alpha^4$ and $m\alpha^4(m/M)$ (including only terms that are required for calculation of spin-averaged energies) is given by the sum:
\begin{equation}
H^{(4)} = H_B + H_{ret} + H_{S},
\end{equation}
where
\begin{equation} \label{breit-pauli}
\begin{array}{@{}l}\displaystyle
H_B = -\frac{p_e^4}{8m^3}+\frac{[\Delta_e V_e]}{8m^2},
\\[3mm]\displaystyle
H_{ret} =
   \frac{Z_1}{2}\:
      \frac{p_e^i}{m}\left(\frac{\delta^{ij}}{r_1}+\frac{r_1^ir_1^j}{r_1^3}\right)\frac{P_1^j}{M_1}
   +\frac{Z_2}{2}\:
      \frac{p_e^i}{m}\left(\frac{\delta^{ij}}{r_2}+\frac{r_2^ir_2^j}{r_2^3}\right)\frac{P_2^j}{M_2},
\\[3.5mm]\displaystyle
H_{S} =
   \frac{Z_a}{2m^2}\,\frac{[\mathbf{r}_a\!\times\!\mathbf{p}_e]}{r_a^3}\,\mathbf{s}_e
   -\frac{Z_a}{mM_a}\,\frac{[\mathbf{r}_a\!\times\!\mathbf{P}_a]}{r_a^3}\,\mathbf{s}_e.
\end{array}
\end{equation}

The effective Hamiltonian for corrections of $m\alpha^6$ and $m\alpha^6(m/M)$ orders were derived in \cite{Zhong18,Korobov20}. There are three second-order terms which contribute at these orders to the spin-averaged energy of a bound state. Namely, the leading order nonrecoil contribution
\begin{equation}
E_B = \left\langle
   H_B\,Q (E_0-H_0)^{-1} Q\,H_{B}
\right\rangle,
\end{equation}
where $\langle \rangle$ denotes the expectation value for the state $|\psi_0 \rangle$ under consideration, which is a solution of the three-body Schr\"odinger equation $H_0 \psi_0 = E_0 \psi_0$; $Q$ is a projection operator on a subspace orthogonal to $|\psi_0 \rangle$. The second term is the recoil contribution at the leading order in $m/M$:
\begin{equation}
E_R = 2\left\langle
   H_B\,Q (E_0-H_0)^{-1} Q\,H_{ret}
\right\rangle.
\end{equation}
Both contributions are divergent; the singular part should be separated and will cancel out with that coming from first-order contribution of the $m\alpha^6$-order effective Hamiltonian. The last one is the spin-orbit scalar contribution, which is regular,
\begin{equation} \label{eq:ES0}
E_S^{(0)} =
   \left\langle
      H_{S} Q (E_0-H_0)^{-1} Q H_{S}
   \right\rangle^{(0)}.
\end{equation}
For completeness we also include in our tables the rank 1 contribution of the spin-orbit second-order correction
\begin{equation} \label{eq:ES2}
\Delta H_{S} =
   \left\langle
      H_{S} Q (E_0-H_0)^{-1} Q H_{S}
   \right\rangle^{(1)} =
   E_S^{(1)}\left(\mathbf{L}\cdot\mathbf{s}_e\right),
\end{equation}
which contributes to the spin-orbit interaction coefficient~\cite{Korobov20,Haidar22}.

\section{Separation of singularities}

The second-order contributions $E_B$ and $E_R$ to the spin-averaged energy are singular (divergent) and may be split into regular and singular parts using the following transformation:
\begin{equation}\label{HPP}
{H'_{B\!}}=H_B^{}-(E_0\!-\!H_0)U-U(E_0\!-\!H_0),
\end{equation}
where $U=c_1/r_1+c_2/r_2$ and
\begin{equation}
c_1 = \frac{\mu_1(2\mu_1\!-\!m_e)}{4m_e^3}\>Z_1,
\qquad
c_2 = \frac{\mu_2(2\mu_2\!-\!m_e)}{4m_e^3}\>Z_2.
\end{equation}
Then the second-order term for $E_B$ is transformed as follows:
\begin{equation}
\begin{array}{@{}l}\displaystyle
\left\langle
   H_B\,Q (E_0-H_0)^{-1} Q\,H_B
\right\rangle =
   \left\langle
      H'_B\,Q (E_0\!-\!H_0)^{-1} Q\,H'_B
   \right\rangle
   +\left\langle UH_B\!+\!H_BU \right\rangle
   -2\left\langle U \right\rangle \left\langle H_B \right\rangle
   -\left\langle U(E_0\!-\!H_0)U \right\rangle,
\end{array}
\end{equation}
and the last three terms are added as new interactions to the $m\alpha^6$-order effective Hamiltonian~\cite{Zhong18,Korobov25}. We treat the $E_R$ correction in a similar way. The regular parts of these contributions read:
\begin{equation}
\begin{array}{@{}l}\displaystyle
E'_B =
\left\langle
   H'_B\,Q (E_0-H_0)^{-1} Q\,H'_{B}
\right\rangle, \label{eq:EBprime}
\\[3mm]\displaystyle
E'_R =
2\left\langle
   H'_B\,Q (E_0-H_0)^{-1} Q\,H_{ret}
\right\rangle.
\end{array}
\end{equation}
In the following, we consider in numerical calculations only the second-order corrections in this regularized form.

\section{Numerical approach}

The variational bound state wave functions are calculated by solving the three-body Schr\"{o}dinger equation with Coulomb interaction using the variational approach based on the exponential expansion with randomly chosen exponents~\cite{Korobov00}. Details and particular strategy of choice of the variational nonlinear parameters and basis structure that have been adopted in the present work can be found in~\cite{Korobov20,Haidar22}.

Briefly, the wave function for a state with a total orbital angular momentum $L$ and of a total spatial parity $\pi=(-1)^L$ is expanded as follows:
\begin{equation}\label{exp_main}
\begin{array}{@{}l}
\displaystyle \Psi_{LM}^\pi(\mathbf{R},\mathbf{r}_1) =
       \sum_{l_1+l_2=L}
         \mathcal{Y}^{l_1l_2}_{LM}(\hat{\mathbf{R}},\hat{\mathbf{r}}_1)
         G^{L\pi}_{l_1l_2}(R,r_1,r_2),
\\[4mm]\displaystyle
G_{l_1l_2}^{L\pi}(R,r_1,r_2) =
    \sum_{n=1}^N \Big\{C_n\,\mbox{Re}
          \bigl[e^{-\alpha_n R-\beta_n r_1-\gamma_n r_2}\bigr]
+D_n\,\mbox{Im} \bigl[e^{-\alpha_n R-\beta_n r_1-\gamma_n r_2}\bigr] \Big\},
\end{array}
\end{equation}
where the complex exponents, $\alpha_n$, $\beta_n$, $\gamma_n$, are generated in a pseudorandom way.

Second-order terms, having an expression of the type $AQ(E_0 - H_0)^{-1}QB$, are evaluated by solving numerically the equation
\begin{equation} \label{eq:psi1}
(E_0 - H_0) \psi^{(1)} = \left( B - \langle B \rangle \right) \psi_0,
\end{equation}
and calculating the scalar product $\langle \psi_0 | A | \psi^{(1)} \rangle$. In order to solve Eq.~(\ref{eq:psi1}), $\psi^{(1)}$ is expanded in a variational basis according to Eq.~(\ref{exp_main}). Given that, for $B = H'_B$, the $\psi^{(1)}$ function behaves like $\ln(r_a)$ at small electron-nucleus distances, the basis set must include high exponents $\beta_n$ and $\gamma_n$. We use a ``multi-layered''  basis where the first subsets (between two and
four) approximate the regular part of the intermediate wave function, and others subsets contain growing exponents in order to reproduce the singular behavior at small distances. An illustrative example can be found in Table I of ~\cite{Korobov20}.

The operators $H'_{B}$ and $H_{ret}$ involved in Eq.~(\ref{eq:EBprime}) are scalar, so that $\psi^{(1)}$ has an angular momentum $L'=L$. The spin-orbit operator $H_S$ that appears in the contributions ~(\ref{eq:ES0})-(\ref{eq:ES2}) is of rank 1; for this case, the calculation in divided into the angular momentum components $L'=L-1,L,L+1$. We calculate the quantities:
\begin{equation} \label{components-so-so}
\begin{array}{@{}l}\displaystyle
a_- = -\frac{1}{2L+1}
   \sum_{n\ne0} \frac{\left\langle vL\|\mathbf{A}_{S}\|v_nL-1\right\rangle
   \left\langle v_nL-1\|\mathbf{A}_{S}\|vL-1\right\rangle}{E_0-E_n}\,,
\\[3mm]\displaystyle
a_0 = \frac{1}{2L+1}
   \sum_{n\ne0} \frac{\left\langle vL\|\mathbf{A}_{S}\|v_nL\right\rangle
   \left\langle v_nL\|\mathbf{A}_{S}\|vL\right\rangle}{E_0-E_n}\,,
\\[3mm]\displaystyle
a_+ = -\frac{1}{2L+1}
   \sum_{n\ne0} \frac{\left\langle vL\|\mathbf{A}_{S}\|v_nL+1\right\rangle
   \left\langle v_nL+1\|\mathbf{A}_{S}\|vL+1\right\rangle}{E_0-E_n}\,.
\end{array}
\end{equation}
Here, $\mathbf{A}_{S}$ is the spatial part of the spin-orbit Hamiltonian $H_{S}$ in Eq.~(\ref{breit-pauli}), i.e. $H_{S} = \mathbf{A}_{S} \!\cdot\! \mathbf{s}_e$. The scalar and vector components are obtained from the following formulas:
\begin{equation}
\begin{array}{@{}l}\displaystyle
E_S^{(0)} = \frac{1}{4} \left( a_- + a_0 + a_+ \right)\,,
\\[3mm]\displaystyle
E_S^{(1)} = -\frac{1}{2}\left[\frac{a_-}{L}+\frac{a_0}{L(L\!+\!1)}-\frac{a_+}{L+1}\right].
\end{array}
\end{equation}

\begin{table}
\caption{Convergence of the $E'_B$ second-order energy correction for the ground $(L=0,v=0)$ state of HD$^+$ as a function of the basis size $N$ for the initial state and $N'$ for the intermediate state.}\label{conv}
\begin{center}
\begin{tabular}{cccc}
\hline\hline
 $N\backslash N'$ &    8000  &     10000 &      12000  \\
\hline
 max $\beta$ or $\gamma$ & $10^4$ & $10^5$ & $10^6$ \\
\hline
 3000 & $-$0.3009300 & $-$0.3009349 & $-$0.3019389 \\
 4000 & $-$0.3009292 & $-$0.3009336 & $-$0.3009345 \\
 5000 & $-$0.3009292 & $-$0.3009335 & $-$0.3009340 \\
 6000 & $-$0.3009292 & $-$0.3009335 & $-$0.3009339 \\
\hline\hline
\end{tabular}
\end{center}
\end{table}

\section{Results and Discussion}

In Table~\ref{conv}, we study the convergence of the $E'_B$ term, which is the numerically most challenging one, as a function of the size of the basis sets. The size $N'$ of the ``intermediate'' basis set representing $\psi^{(1)}$ is increased from $N'=8000$ to $N'=12000$, and at the same time higher exponents are incorporated, with a maximal exponent value growing from $10^4$ to $10^6$. We also study the effect of the size $N$ of the basis set that represents the initial state $\psi_0$; it is found that relatively large values of $N$ are required for good convergence. This is due to the singular behavior discussed above, which requires a highly accurate approximation of the wave function for small $r_1$ and $r_2$. The results of Table~\ref{conv} provide us with necessary guidelines regarding how to choose properly both the initial and intermediate state parameters.

Our final numerical results for the H$_2^+$ and HD$^+$ molecular ions in a range of rotational and vibrational states: $L=0-4$, $v=0-4$, are presented in Tables \ref{H2plus-av} and \ref{HDplus-av}. We expect that all the digits indicated in the Tables are converged. The uncertainty of ro-vibrational transition frequencies originating from these terms, which is dominated by the largest one $E'_B$, is therefore smaller than $10^{-6} \alpha^4 E_h/h \simeq 19$~Hz, which is compatible with a precision improvement by 1-2 orders of magnitude depending on the transition.

\begin{table}[t]
\caption{Contribution of various second-order terms to the energies of the ro-vibrational states $(L,v)$ in the H$_2^+$ molecular ion. $a[b]=a\times10^b$.}
\label{H2plus-av}
\begin{center}
\begin{tabular}{c@{\hspace{3mm}}r@{\hspace{8mm}}c@{\hspace{8mm}}c@{\hspace{8mm}}c@{\hspace{8mm}}c}
\hline\hline
\vrule width0pt height 11pt
 & $v$ & $E'_B$ & $E'_R$ & $E_S^{(0)}$ & $E_S^{(1)}$ \\
\hline
\vrule width0pt height 12pt
$L=0$& 0& $-$0.299920   & 0.158162[$-$02] & $-$0.854125[$-$02] & 0  \\
  & 1   & $-$0.291126   & 0.153454[$-$02] & $-$0.794757[$-$02] & 0  \\
  & 2   & $-$0.283222   & 0.149254[$-$02] & $-$0.738831[$-$02] & 0  \\
  & 3   & $-$0.276144   & 0.145526[$-$02] & $-$0.685928[$-$02] & 0  \\
  & 4   & $-$0.269841   & 0.142242[$-$02] & $-$0.635664[$-$02] & 0  \\ [1.5mm]
$L=1$& 0& $-$0.299427   & 0.157916[$-$02] & $-$0.851331[$-$02] & 0.22326[$-$04] \\
  & 1   & $-$0.290672   & 0.153228[$-$02] & $-$0.792132[$-$02] & 0.21066[$-$04] \\
  & 2   & $-$0.282803   & 0.149047[$-$02] & $-$0.736361[$-$02] & 0.19847[$-$04] \\
  & 3   & $-$0.275759   & 0.145337[$-$02] & $-$0.683600[$-$02] & 0.18659[$-$04] \\
  & 4   & $-$0.269490   & 0.142071[$-$02] & $-$0.633466[$-$02] & 0.17494[$-$04] \\[1.5mm]
$L=2$& 0& $-$0.298451   & 0.157428[$-$02] & $-$0.845783[$-$02] & 0.22164[$-$04] \\
  & 1   & $-$0.289771   & 0.152781[$-$02] & $-$0.786921[$-$02] & 0.20913[$-$04] \\
  & 2   & $-$0.281973   & 0.148637[$-$02] & $-$0.731459[$-$02] & 0.19701[$-$04] \\
  & 3   & $-$0.274997   & 0.144963[$-$02] & $-$0.678981[$-$02] & 0.18525[$-$04] \\
  & 4   & $-$0.268792   & 0.141730[$-$02] & $-$0.629108[$-$02] & 0.17374[$-$04] \\[1.5mm]
$L=3$& 0& $-$0.297009   & 0.156708[$-$02] & $-$0.837565[$-$02] & 0.21921[$-$04] \\
  & 1   & $-$0.288441   & 0.152120[$-$02] & $-$0.779201[$-$02] & 0.20681[$-$04] \\
  & 2   & $-$0.280749   & 0.148033[$-$02] & $-$0.724195[$-$02] & 0.19480[$-$04] \\
  & 3   & $-$0.273873   & 0.144411[$-$02] & $-$0.672137[$-$02] & 0.18313[$-$04] \\
  & 4   & $-$0.267764   & 0.141229[$-$02] & $-$0.622648[$-$02] & 0.17175[$-$04] \\
\hline\hline
\end{tabular}
\end{center}
\end{table}

\begin{table}[t]
\caption{Similar to Table~\ref{H2plus-av}, for the HD$^+$ molecular ion. $a[b]=a\times10^b$.}
\label{HDplus-av}
\begin{center}
\begin{tabular}{c@{\hspace{3mm}}r@{\hspace{8mm}}c@{\hspace{8mm}}c@{\hspace{8mm}}c@{\hspace{8mm}}c}
\hline\hline
\vrule width0pt height 11pt
 & $v$ & $E'_B$ & $E'_R$ & $E_S^{(0)}$ & $E_S^{(1)}$ \\
\hline
\vrule width0pt height 12pt
$L=0$& 0& $-$0.300935   & 0.118927[$-$02] & $-$0.858582[$-$02] & 0  \\
  & 1   & $-$0.293200   & 0.115782[$-$02] & $-$0.806713[$-$02] & 0  \\
  & 2   & $-$0.286143   & 0.112933[$-$02] & $-$0.757492[$-$02] & 0  \\
  & 3   & $-$0.279723   & 0.110360[$-$02] & $-$0.710643[$-$02] & 0  \\
  & 4   & $-$0.273902   & 0.108048[$-$02] & $-$0.665907[$-$02] & 0  \\[1.5mm]
$L=1$& 0& $-$0.300562   & 0.118785[$-$02] & $-$0.856472[$-$02] & 0.16821[$-$04] \\
  & 1   & $-$0.292852   & 0.115651[$-$02] & $-$0.804714[$-$02] & 0.15998[$-$04] \\
  & 2   & $-$0.285820   & 0.112811[$-$02] & $-$0.755598[$-$02] & 0.15196[$-$04] \\
  & 3   & $-$0.279422   & 0.110247[$-$02] & $-$0.708844[$-$02] & 0.14408[$-$04] \\
  & 4   & $-$0.273625   & 0.107944[$-$02] & $-$0.664191[$-$02] & 0.13600[$-$04] \\
  & 5   & $-$0.268417   & 0.105889[$-$02] & $-$0.621389[$-$02] & 0.12662[$-$04] \\[1.5mm]
$L=2$& 0& $-$0.299822   & 0.118504[$-$02] & $-$0.852277[$-$02] & 0.16730[$-$04] \\
  & 1   & $-$0.292162   & 0.115389[$-$02] & $-$0.800741[$-$02] & 0.15911[$-$04] \\
  & 2   & $-$0.285176   & 0.112569[$-$02] & $-$0.751830[$-$02] & 0.15115[$-$04] \\
  & 3   & $-$0.278823   & 0.110023[$-$02] & $-$0.705267[$-$02] & 0.14337[$-$04] \\
  & 4   & $-$0.273068   & 0.107737[$-$02] & $-$0.660792[$-$02] & 0.13565[$-$04] \\[1.5mm]
$L=3$& 0& $-$0.298725   & 0.118087[$-$02] & $-$0.846042[$-$02] & 0.16593[$-$04] \\
  & 1   & $-$0.291140   & 0.115002[$-$02] & $-$0.794837[$-$02] & 0.15779[$-$04] \\
  & 2   & $-$0.284224   & 0.112210[$-$02] & $-$0.746232[$-$02] & 0.14989[$-$04] \\
  & 3   & $-$0.277938   & 0.109691[$-$02] & $-$0.699953[$-$02] & 0.14219[$-$04] \\
  & 4   & $-$0.272247   & 0.107431[$-$02] & $-$0.655743[$-$02] & 0.13462[$-$04] \\
  & 9   & $-$0.251790   & 0.099615[$-$02] & $-$0.457845[$-$02] & 0.09850[$-$04] \\
\hline\hline
\end{tabular}
\end{center}
\end{table}

These results should be combined with calculation of the first-order contributions from the effective Hamiltonians of order $m\alpha^6$ and $m\alpha^6(m/M)$ (modified by addition of the singular terms that have been subtracted from the second-order contributions in the present work) in order to get the total energy corrections at these orders. Regularization of these Hamiltonians, leading to expressions in terms of finite expectation values, was recently performed~\cite{Korobov25,Korobov25b}, and numerical calculations are currently in progress.

\section{Acknowledgements}

Z.-X.Z. acknowledges support by the National Natural Science Foundation of China under Grant No. 12393821.

\end{document}